\documentclass[11pt,peerreview,letterpaper]{IEEEtran}
\usepackage{textcomp}
\usepackage{psfrag}
\usepackage{graphicx}  
\usepackage{graphics}

\usepackage{url}
\usepackage[nospace]{cite}
\usepackage{color}
\usepackage[all]{xy}
\usepackage{amsmath,amsfonts,amssymb}
\usepackage{mathrsfs}
\usepackage[hang,normalsize]{subfigure}
\usepackage{psfrag}
\usepackage{epstopdf}
\usepackage{times}

\newtheorem{theorem}{Theorem}[section]
\newtheorem{corollary}{Corollary}[section]
\newtheorem{prop}[theorem]{Proposition}
\newtheorem{lemma}[theorem]{Lemma}

\newtheorem{definition}{Definition}[section]
\newtheorem{conjecture}{Conjecture}[section]
\newtheorem{example}{Example}[section]

\renewcommand{\bar}{\overline}

\newcommand{\iprod}[2]{\langle #1,#2\rangle}

\renewcommand{\to}{\rightarrow}

\renewcommand{\span}{\mathrm{span}}
\newcommand{\spe}{\mathrm{span}_\epsilon}

\newcommand{\bE}{{\bf{E}}}

\newcommand{\bH}{{\bf{H}}}

\newcommand{\bJ}{{\bf{J}}}

\newcommand{\br}{{\bf{r}}}

\newcommand{\bu}{{\bf{u}}}
\newcommand{\bv}{{\bf{v}}}

\newcommand{\bx}{{\bf{x}}}
\newcommand{\by}{{\bf{y}}}

\newcommand{\cL}{\mathcal{L}}

\newcommand{\cN}{\mathcal{N}}

\newcommand{\cH}{\mathcal{H}}
\renewcommand{\cN}{\mathcal{N}}

\newcommand{\bbR}{\mathbb{R}}

\newcommand{\bbZ}{\mathbb{Z}}
\newcommand{\bbC}{\mathbb{C}}

\graphicspath{{./fig/}}

\begin{document}
\title{Degrees of Freedom of a Communication Channel: Using Generalised Singular Values}
\bibliographystyle{IEEEtran}

\author{Ram Somaraju and Jochen Trumpf%
%\thanks{Ram Somaraju and Jochen Trumpf are with the Department of Information Engineering,
%Research School of Information Sciences and Engineering, Building 115,
%The Australian National University,
%Canberra, ACT 0200,
%Australia and National ICT Australia Limited, Locked Bag 8001,
%Canberra, ACT 2601,
%Australia.}% <-this % stops a space
%\thanks{National ICT Australia Limited is funded by the Australian
%Government's Department of Communications, Information Technology
%and the Arts and the Australian Research Council through Backing
%Australia's Ability and the ICT Centre of Excellence Program.}
}

\markboth{}{}
\maketitle

\begin{abstract}
A fundamental problem in any communication system is: given a
communication channel between a transmitter and a receiver, how many
``independent'' signals can be exchanged between them? Arbitrary
communication channels that can be described by linear compact
channel operators mapping between normed spaces are examined in this
paper. The (well-known) notions of degrees of freedom at
level~$\epsilon$ and essential dimension of such channels are
developed in this general setting. We argue that the degrees of
freedom at level~$\epsilon$ and the essential dimension fundamentally
limit the number of independent signals that can be exchanged between
the transmitter and the receiver. We also generalise the concept of singular values of
compact operators to be applicable to compact operators
defined on arbitrary normed spaces which do not necessarily carry a
Hilbert space structure. We show how these generalised singular values can be used
to calculate the degrees of freedom at level~$\epsilon$ and the
essential dimension of compact operators that describe communication
channels. We describe physically realistic channels that require such
general channel models.

\end{abstract}


\IEEEpeerreviewmaketitle

\section{Introduction}\label{int}
The basic consideration in this paper can be stated as follows: given an arbitrary communication channel, is it possible to evaluate the number of independent sub-channels or modes available for communication. Though this question is not generally examined explicitly, it plays an important role in various information theoretic problems. 

A rigorous proof of Shannon's famous capacity
result~\cite{Shannon1948} for continuous-time band-limited white
Gaussian noise channels requires a calculation of the number of approximately time-limited and band-limited sub-channels (see e.g.~\cite[ch. 8]{Gallagher1968} and~\cite{Verdu1998,Landau1962}). This result can be generalised to dispersive/non-white Gaussian channels using the \emph{water-filling} formula~\cite{Shannon1948,Gallagher1968}. In order to use this formula, one needs to diagonalise the channel operator and allocate power to the different sub-channels or modes based on the singular values of the corresponding sub-channel. One therefore needs to calculate the modes and the power transferred (square of the singular values) on each one of these sub-channels to calculate the channel capacity.

The water-filling formula has been used extensively in order to
calculate the capacity of channels that use different forms of
diversity. In particular, the capacity of multiple-input
multiple-output (MIMO) antenna systems has been calculated using this
water-filling formula for various conditions imposed on the
transmitting and the receiving antennas (see e.g.~\cite{Biglieri2004}
and references therein). Water-filling type formulas have been used
for other multi-access schemes such as OFDM-MIMO~\cite{Bolcskei2002}
and CDMA~\cite{Grant1998} (see also Tulino~\cite[sec 1.2]{Tulino2004} and references therein). More recently, several papers have examined the number of degrees of freedom\footnote{Note that other terms such as \emph{modes of communication}, \emph{essential dimension} etc. have been used instead of degrees of freedom in some of these papers.} available in spatial channels~\cite{Poon2005,Hanlen2006,Kennedy2007,Xu2006,Migliore2006}. Questions of this nature have also been studied in other contexts such as optics~\cite{Miller2000} and spatial sampling of electromagnetic waves~\cite{Bucci1987,Bucci1989}. 

Both types of results, the modes of communication used for the
water-filling formula and the number of degrees of freedom of spatial
channels use the singular value decomposition (SVD) theorem. One can
use SVD to diagonalise the channel operator and the magnitude of the
singular values determines the power transferred on each of the
sub-channels. The magnitude of these singular values can therefore be
used to calculate the number of degrees of freedom of the channel (see
e.g.~\cite{Xu2006,Poon2005}). However, the SVD theorem is only
applicable to compact operators defined on Hilbert spaces. An implicit
and valid assumption that is used in these papers is that the
operators describing the communication channels are defined on Hilbert
spaces. These results can therefore not be generalised directly to
communication systems that are modeled by operators defined on normed
spaces that do not admit an inner product structure. There are several
instances of practical channels that can not be modeled using
operators defined on inner-product spaces (see Section~\ref{egs} for
examples). In this paper, we develop a general theory that enables one to evaluate the number of degrees of freedom of such systems. 

We wish to examine if it is possible to evaluate the number of
parallel sub-channels available in general communication systems that
can be described using linear compact operators. Any communication
channel is subject to various physical constraints such as noise at
the receiver or finite power available for transmission. If the
channel can be modeled via a linear compact operator, then these
constraints ensure that only finitely many independent channels are
available for communication. Roughly speaking, we call the number of
such channels the number of degrees of freedom of the communication
system (see Section~\ref{maiRes} for a precise definition). Note that
if the channel is modeled using a linear operator that is not
compact then it will in fact have infinitely many parallel
sub-channels, or some channels that can transfer an infinite amount of
power (see Theorem~\ref{Nthe:comp} below and the discussion following
it). It could hence be argued that the theory presented in this paper
is the most general theory needed to model physically realistic channels.

We give novel definitions for the terms degrees of freedom and
essential dimension in the following section. Even though these terms
have been used interchangeably in the literature, we distinguish
between the two. The essential dimension of a channel is useful for
channels that have numbers of degrees of freedom that are essentially
independent of the receiver noise level (e.g. the
time-width/band-width limited channels in Slepian's work~\cite{Slepian1976}). Also, we generalise the notion of singular values to compact operators defined on normed spaces and explain how these generalised singular values can be used to compute degrees of freedom and the essential dimension. 

\subsection{Channel Model} \label{chaMod}
We assume that a communication channel between a transmitter and a
receiver can be modeled as follows. Let $X$ be a linear vector space
of functions that the transmitter can generate and let $Y$ be a linear
vector space of functions that the receiver can measure. We assume the
existence of a linear operator $T:X\to Y$ that maps each signal
generated by a transmitter to a signal that a receiver can measure. We
also assume that there is a norm $\|\cdot\|_X$ on $X$ and a norm
$\|\cdot\|_Y$ on $Y$. This model is very general and can be applied to
various situations of practical relevance. 

For instance, consider a MIMO communication system wherein the
transmitter symbol waveform shape on each antenna is a raised
cosine. In this case we can think of the space of transmitter
functions $X$ to be (more precisely, to be parametrised by) the
$n$-dimensional complex space $\bbC^n$ that 
determines the phase and amplitude of the raised cosine waveform on
each antenna. Here $n$ is the number of transmitting antennas. Also,
we can think of the space of receiver functions as $\bbC^m$, where $m$
is the number of receiving antennas. $T$ in this context is a
channel matrix, representing the linearized channel operator that
depends on the scatterers in the environment.  

Alternatively, consider a MIMO communication system in which the transmitter symbols are not fixed but can be any waveform of time. Suppose the symbol time is fixed to $t_{s}$ seconds. In this case, we can think of the space of transmitter functions, $X$, as the space $\cL^2([0,t_{s}],\bbC^n)$ of $\bbC^n$-valued square integrable functions defined on $[0,t_{s}]$. Similarly, we can think of the space of receiver functions, $Y$, as the space $\cL^2([0,t_{s}],\bbC^m)$. Again, $T$ is the channel operator. 

Irrespective of the precise form of the underlying spaces $X$ and $Y$, we always call elements of $X$ transmitter functions and the elements of $Y$ receiver functions. Also, we call the space $X$ the space of transmitter functions and the space $Y$ the space of receiver functions. In particular, we do not distinguish between the two different physical situations: a) the elements of $X$ are functions of time and b) the elements of $X$ are vectors in some finite dimensional space. This should cause no confusion and we use this convention for the remainder of this document. 

We now restrict ourselves to situations where there is a source
constraint $\|\cdot\|_X \leq P$ that can be imposed on the space of
transmitter functions $X$, and where the operator $T$ is
compact. Roughly speaking, the norm on the space of transmitter
functions $X$ captures the physical restriction that the transmitter
functions can not be \emph{arbitrarily big}, while the norm on the space of receiver functions can be interpreted as a measure of how \emph{big} the received signals are compared to a pre-specified noise level. We therefore try to find how many linearly independent signals can be generated at the receiver that are \emph{big} enough by transmitter functions that are not \emph{too big}. The compactness of the operator $T$ ensures that only finitely many independent signals can be received (see Section~\ref{egs} for examples of such channels). This vague idea is clarified further in the following two sections. 

\subsection{Outline} The remainder of this paper is organised as
follows: in the next section we consider a finite dimensional example
and motivate the definition of degrees of freedom. We also discuss
several examples of practical communication systems to which the
theory developed in this paper may be applied.  Section~\ref{maiRes}
presents the main results of this paper as well as formal definitions
of degrees of freedom, essential dimension and generalised singular
values. Conclusions are presented in Section~\ref{chaCon}. Detailed
proofs of the theorems in this paper are presented in the Appendix. 

Most of the material presented in this paper forms part of the first
author's PhD thesis~\cite{thesis}.
%%% Local Variables: 
%%% mode: latex
%%% TeX-master: "main"
%%% End: 

\section{Motivation}\label{EDMot} We motivate our definition of degrees of freedom at level~$\epsilon$ for compact operators on normed spaces by considering linear operators on finite dimensional spaces. Consider a communication channel that uses $n$ transmitting antennas and $m$ receiving antennas which can be mathematically modeled as follows. Let the current on the $n$ transmitting antennas be given by $\bx\in\bbC^n$. This current on the transmitting antennas generates a current $\by \in \bbC^m$ in the $m$ receiving antennas according to the equation
\begin{equation*}
	\by = \bH\bx.
\end{equation*}
Here, $\bH\in\bbC^{m\times n}$ is the channel matrix. We can define the operator $T:\bbC^n\to\bbC^m$ by $\bx\mapsto\by=\bH\bx$. Also, for  $n = 1,2,\ldots$, $\|\cdot\| = \sqrt{(\cdot)^\ast (\cdot)}$, with $(\cdot)^\ast$ denoting the complex conjugate transpose, is the standard norm in $\bbC^n$. In this context, the norm determines the power of the signal on the antennas. 

The singular value decomposition theorem tells us that there exist sets of orthonormal basis vectors $\{\bv_1,\ldots,\bv_n\}\subset\bbC^n$ and $\{\bu_1,\ldots,\bu_m\}\subset\bbC^m$ such that the matrix representation for $T$ in these bases is diagonal. Let $\bH_{d}$ be such a matrix with the basis vectors ordered such that the diagonal elements (i.e. the singular values of $T$) are in non-increasing order. A simple examination of the diagonal matrix proves that
% \begin{enumerate}
% 	\item for all $\epsilon>0$ there exist a number $N_1$ and a set of mutually orthogonal vectors $\{\bv_1,\ldots,\bv_{N_1}\} \subset\bbC^n$ such that if any vector $\bv,\ \|\bv\|\leq 1$ is orthogonal to $\bv_i, i=1,\ldots,N_1$, then $\|\bH\bv\|<\epsilon$. Here $\|\cdot\|$ denotes the standard norm on $\bbC^n$. For a given $\epsilon$, call the smallest number that satisfies the above condition $N_1(\epsilon)$. Note that the vectors ${\bv_1,\ldots,\bv_{N_1}}$ span the space of all linear combinations of the right singular vectors of $H$ whose corresponding singular values are greater than or equal to $\epsilon$. 
%	\item 
for all $\epsilon>0$ there exist a number $N$ and a set of linearly independent vectors $\{\by_1,\ldots,\by_{N}\}\subset\bbC^m$ such that for all $\bx\in \bar{B}_{1,\bbC^n}(0)$\footnote{Given a normed space $X$, $r \geq 0$ and $x\in X$, $\bar{B}_{r,X}(x)$ denotes the closed ball of radius $r$ centered at $x\in X$.} 
\begin{equation*}
	\inf_{a_1,\ldots,a_N}\left\|\bH_{d}\bx-\sum_{i=1}^{N} a_i\by_i\right\| \leq \epsilon.
\end{equation*}
For a given $\epsilon$, call the smallest number that satisfies the above condition $\cN(\epsilon)$. Note that the vectors ${\by_1,\ldots,\by_{N}}$ span the space of all linear combinations of the left singular vectors of $T$ whose corresponding singular values are greater than or equal to $\epsilon$.  
%\end{enumerate}

A simple examination of the diagonal matrix tells us that 
%both $N_1(\epsilon)$ and 
$\cN(\epsilon)$ is equal to the number of singular values of $T$ that
are greater than $\epsilon$ and is hence clearly independent of the
bases chosen. This leads us to our definition for degrees of freedom
in finite dimensional spaces. 
\begin{definition}\label{defFin} Let $T:\bbC^n\to\bbC^m$ be a linear operator and let $\epsilon> 0$ be given. Then the number of degrees of freedom at level~$\epsilon$ for $T$ is the smallest number $N$ such that there exists a set of vectors $\by_1,\ldots,\by_N\in\bbC^m$ such that  for all $\bx\in \bar{B}_{1,\bbC^n}(0)$
\begin{equation*}
	\inf_{a_1,\ldots,a_N} \left\|T\bx-\sum_{i=1}^Na_i\by_i\right\| \leq \epsilon.
\end{equation*}
\end{definition}
This definition is appropriate for the number of degrees of freedom
because for a MIMO system the norm $\|\cdot\|$ represents the power in
the signal. Suppose we wish to transmit $N$ linearly independent
signals from the transmitter to the receiver, and the total power
available for transmission is bounded. Suppose further that the
received signal is measured in the presence of noise. By requiring
that $\bx\in \bar{B}_{1,\bbC^n}(0)$ we are constraining the power
available for transmission. We model the noise by assuming that any
two signals at the receiver can be distinguished if the power of the
difference between the signals is greater than some
level~$\epsilon$. Similar ideas have been used for instance by 
Bucci~\emph{et. al.}~\cite{Bucci1989} (see
also~\cite{Hanlen2006,Landau1962,Slepian1976}). According to this
definition, the number of degrees of freedom is equal to the number of
linearly independent signals that the receiver can distinguish under
the assumptions of a transmit power constraint and a receiver noise
level represented by $\epsilon$. Note that we are making the implicit
assumption that the power $P$ is 1 in the above definition. This does
not cause a problem because we can always scale the norm in order to consider situations where $P\neq 1$.

The above definition was motivated using the singular value
decomposition theorem in finite dimensional spaces. It can therefore
be easily generalised to infinite dimensional Hilbert spaces using the
corresponding singular value decomposition in infinite dimensional
Hilbert spaces (see  eg.~\cite{Bucci1989,thesis}\footnote{Also
  compare with the time-bandwidth problem
  in~\cite{Landau1962,Slepian1976}.}). However, the singular value
decomposition can only be used for operators defined on Hilbert
spaces. It cannot be used for operators defined on general normed
spaces. Observe that the definition for degrees of freedom above only depends on the norm $\|\cdot\|$ and not on the assumption that the underlying spaces $\bbC^n$ and $\bbC^m$ are Hilbert spaces. It will be shown in this paper that the above definition can be extended to compact operators defined on arbitrary normed spaces. 

Now consider the situation where the singular values of the operator $T$ show a step like behavior. For instance, suppose the singular values are $\{1,0.9,0.85,0.5,0.1,0.05,.0005\}$. In this particular case the number of degrees of freedom at level~$\epsilon$ is equal to $4$ for a big range of values of $\epsilon$ and the number of degrees of freedom is essentially independent of the actual value of $\epsilon$ chosen. Such a situation arises in several important cases (see eg.~\cite{Miller2000,Slepian1976,Landau1962,Bucci1989,Poon2005}). It would be useful to have a general way in which one can specify a number of degrees of freedom of a channel that is independent of the arbitrarily chosen level~$\epsilon$. In this paper we provide a novel definition for such a number and call it the essential dimension of the channel. This definition is sufficiently general to be applicable to a variety of channels and quantifies the essential dimension of any channel that can be described using a compact operator. 

\subsection{Examples}\label{egs} As explained in section~\ref{chaMod}, we assume that a communication channel can be described using the triple $X$, $Y$ and $T$. Here $X$ is the space of transmitter functions, $Y$ is the space of receiver functions and $T$ is the channel operator and is assumed to be compact. As explained earlier in this section, if the spaces $X$ and $Y$ are Hilbert spaces and if the operator $T$ is a linear compact operator then the well known theory of singular values of Hilbert space operators can be used to determine the number of degrees of freedom of such channels. However, if either one of the spaces $X$ or $Y$ is not an inner product space then one cannot use this theory.

There are several practical channels that are best described using abstract spaces that do not admit an inner product structure. In this subsection, we consider three examples of such channels. In the first example, the measurement technique used in the receiver restricts the space of receiver functions. In the second one, the modulation technique used means that the constraints on the space of transmitter functions are best described using a norm that is not compatible with an inner product. The final example discusses a physical channel that naturally admits a norm on the space of transmitter functions that is described using a vector product and therefore does not admit an inner-product structure. 

\begin{example} In any practical digital communication system, the
  receiver is designed to receive a finite set of transmitted
  signals. Suppose the transmitted signal is generated from a source
  alphabet $\{t_1,\ldots,t_N\}$ and for simplicity assume that in a
  noiseless system each element from the source alphabet $t_i, 1\leq i
  \leq N$, generates a signal $r_i, 1\leq i\leq N$, at the
  receiver. In the corresponding noisy system, the fundamental problem
  is to determine which element from the source alphabet was
  transmitted given the signal $r = r_i + n$ was received. Here, $n$
  is the noise in the system. One common approach to solving this
  problem is to define some metric $d(\cdot,\cdot)$ that measures the
  distance between two receiver signals and to calculate 
\begin{equation*} 
r' = \mathop{\mathrm{argmin}}\limits_{\{r_{i},1\leq i \leq n\}} d(r,r_i).
\end{equation*}
One concludes that the element from the source alphabet that
corresponds to $r'$ is (most likely) the transmitted signal. Generally, this metric $d(\cdot,\cdot)$ determines the abstract space $Y$ of receiver function. 
 
Now consider a MIMO antenna system with $n$ transmitting and $m$ receiving antennas. Suppose that the receiver measures the signals on the $m$ receiving antennas for a period of $\tau$ seconds. One can describe the received signal by a function $\by(t)$, where $\by:[0,\tau]\to \bbC^m$. In order to implement the receiver one can use a matched filter if the shapes of all noiseless receiver signals are known. In this case the distance between two received signals can be described using the metric
\begin{equation*}
d(\by_{1},\by_{2}) = \left( \int_0^{\tau} (\by_{1}(t)-\by_{2}(t))^\ast(\by_{1}(t)-\by_{2}(t))dt \right)^{1/2}
\end{equation*}
One can describe the space of receiver functions using the Hilbert space
$\cL^2([0,\tau],\bbC^m)$ with the inner product defined by
\begin{equation*}
\iprod{\by_{1}}{\by_{2}} := \int_0^{\tau} \by_{1}^\ast(t)\by_{2}(t) dt.
\end{equation*}
This is the common approach used in information theory. 

However, it is generally easier to measure just the amplitude of the received signal on each of the $m$ antennas. In fact, in a rapidly changing environment it might not be possible to build an effective matched filter and therefore there is no benefit in measuring the square of the received signal. In this case the distance between any two signals can be described using the metric 
\begin{equation*}
d(\by_{1},\by_{2}) = \int_0^{\tau} |\by_{1}(t)-\by_{2}(t)| dt.
\end{equation*}
Here, one can describe the space of receiver functions using the
Banach space $\cL^1([0,\tau],\bbC^m)$ with the norm defined by
\begin{equation*}
\|\by\| := \int_0^{\tau} |\by(t)| dt.
\end{equation*} 
This channel therefore is best described using a normed space as opposed to an inner product space to model the set of receiver signals. 
\end{example}
\begin{example} Consider a multi-carrier communication system that
  uses some form of amplitude or angle modulation to transmit
  information. Suppose that there are $n$ carriers and that the vector
  $\phi = [\phi_1,\ldots,\phi_n]$ determines the modulating signal on
  each of the carriers. We can think of the modulating waveforms as
  the space of transmitter functions $X$\footnote{In this case we do not
    consider the actual signal on the transmitting antenna (i.e. carrier +
    modulation) to be the transmitter function. Cf. the discussion in Subsection~\ref{chaMod}.}.

If amplitude modulation is used then the vector $\phi$ determines the total power used for modulation. If the total power available for transmission is bounded then one might have an inequality of the form
\begin{equation*}
\sum_{i=1}^n |\phi_i|^2 \leq P.
\end{equation*}
We can therefore describe the space of transmitter functions using the standard Euclidian space $\bbR^n$ with inner product
\begin{equation*}
\iprod{\bx_{1}}{\bx_{2}} = \bx_{1}^T\bx_{2}.
\end{equation*}

Now consider the case where angle modulation is used. In this case all the transmitted signals have the same power and the total power available for transmission places no restrictions on the space of transmitter functions. However, the space of transmitter functions can be subjected to other forms of constraints. For instance, if frequency modulation is used then the maximum frequency deviation used might be bounded by some number $b$ to minimise co-channel interference (see e.g. ~\cite[p. 110,513]{Haykin2000}). Similarly if phase modulation is used the maximum phase variation has to be less than $\pm\pi$. This bound may also depend on other practical considerations such as linearity of the modulator. In this case one might constrain the space of transmitter functions as 
\begin{equation*} 
\sup_{1\leq i\leq n} |\phi_i| < b.
\end{equation*}
The space of transmitter functions of this channel is best described using the $n$-dimensional Banach space $\bbR^n_\infty$ with norm 
\begin{equation*}
\|\bx\| = \sup_{1\leq i\leq n} |\bx_i|.
\end{equation*}
\end{example}
\begin{example}\label{eg:swc} In this final example we examine spatial waveform channels (SWCs)~\cite{thesis}. In SWCs we assume that a current flows in a volume in space and generates an electromagnetic field in a receiver volume that is measured~\cite{Bucci1987,Bucci1989,Hanlen2006,thesis}. Such channels have been used to model MIMO systems previously~\cite{Bucci1987,Bucci1989,Hanlen2006,thesis,Xu2006,Migliore2006}. If a current flows in a volume in space that has a finite conductivity, power is lost from the transmitting volume in two forms. Firstly, power is lost as heat and secondly power is radiated as electromagnetic energy. So the total power lost can be described using the set of equations
\begin{eqnarray*}
P_{total} &=& P_{rad} + P_{lost} \\
P_{lost} &=& \int_V \bJ^*(\br)\bJ(\br) d\br\\
P_{rad} &=& \int_\Omega \bE^*(\br)\times \bH(\br) d\Omega
\end{eqnarray*}
Here, $V$ is some volume that contains the transmitting antennas, $\bJ$ is the current density in the volume $V$ and $\Omega$ is
some sufficiently smooth surface the interior of which contains $V$
with $d\Omega$ denoting a surface area element. Also $\bE$ and $\bH$
are the electric and magnetic fields generated by the current density $\bJ$ and $\cdot \times\cdot$ denotes the vector product in $\bbR^3$. 

Because of the vector product in the last equation above, the total
power lost defines a norm on the space of square-integrable functions
that does not admit an inner-product
structure~\cite{thesis}. The theory developed in this paper is
used to calculate the degrees of freedom of such spatial waveform
channels in~\cite{thesis}.
\end{example}
%%% Local Variables: 
%%% mode: latex
%%% TeX-master: "main"
%%% End: 

\section{Main Results}\label{maiRes} In this section we outline the main results of this paper. All the proofs of theorems are given in the Appendix.
\subsection{Degrees of Freedom for Compact Operators}\label{sec:degr-freed-comp} The definition of degrees of freedom at level~$\epsilon$ for compact operators on normed spaces is identical to the finite dimensional counterpart (Definition~\ref{defFin}) discussed in the previous section with $\bbC^n$ and $\bbC^m$ replaced by general normed spaces. The following theorem ensures that the definition makes sense even in the infinite dimensional setting. 
\begin{theorem}\label{Nthe:defBan}
Suppose $ X$ and $ Y$ are normed spaces with norms $\|\cdot\|_{ X}$
and $\|\cdot\|_{ Y}$, respectively, and $T: X\to  Y$ is a compact
operator. Then for all $\epsilon > 0$ there exist\footnote{$\bbZ$,
  $\bbZ_0^+$ and $\bbZ^+$ are respectively the sets of integers,
  non-negative integers and positive integers.} $N\in \bbZ_0^+$ and a
set $\{\psi_i\}_{i=1}^N\subset  Y$ such that for all $x\in \bar{B}_{1,X}(0)$
\begin{equation*}
    \inf_{a_1,\ldots,a_N} \left\|Tx-\sum_{i=1}^N a_i\psi_i\right\|_{ Y}
    \leq \epsilon .
\end{equation*}
\end{theorem}
Note that for $N=0$ the set $\{\psi_i\}_{i=1}^N$ is empty and the sum
in the above expression is void.
We will use the following definition for the number of degrees
of freedom at level~$\epsilon$ for compact operators on normed spaces.
\begin{definition}[Degrees of freedom at
level~$\epsilon$]\label{Ndef:banDof} Suppose $ X$ and $ Y$
are normed spaces with norms $\|\cdot\|_{ X}$ and $\|\cdot\|_{ Y}$, respectively, and $T: X\to  Y$ is a compact operator.
Then the number of degrees of freedom of $T$ at level~$\epsilon$ is the
smallest $N\in \bbZ_0^+$ such that there exists a set of vectors
$\{\psi_1,\ldots,\psi_N\}\subset Y$ such that for all $x\in \bar{B}_{1,X}(0)$
\begin{equation*}
    \inf_{a_1,\ldots,a_N} \left\|Tx-\sum_{i=1}^N a_i\psi_i\right\|_{ Y}
    \leq \epsilon.
\end{equation*}
\end{definition}
This definition has exactly the same interpretation as in the finite
dimensional case: if there is some constraint $\|\cdot\|_X\leq 1$ on
the space of source functions and if the receiver can only measure
signals that satisfy $\|\cdot\|_Y > \epsilon$, then the number of
degrees of freedom is the maximum number of linearly independent
signals that the receiver can measure under these constraints.   

This definition however is a descriptive one and can not be used to
calculate the number of degrees of freedom for a given compact
operator because the proof of Theorem~\ref{Nthe:defBan} is not
constructive. In the finite dimensional case we can calculate the
degrees of freedom by calculating the singular values. However, as far
as we are aware, there is no known generalisation of singular values
for compact operators on arbitrary normed spaces\footnote{A
  generalisation to compact operators on Hilbert spaces is of course
  classical and well known.}. In the following subsection we will
propose such a generalisation. In fact, we will use the degrees of
freedom to generalise the concept of singular values. 
We will discuss the problem of computing degrees of freedom using
generalised singular values in subsection~\ref{NcomSin} below.

Next, we establish some useful properties of degrees of freedom that
will help motivate the definition of generalised singular values given
in the next subsection.   

\begin{theorem}\label{EDDBthe:dofProp}
Suppose $ X$ and $ Y$ are normed spaces with norms $\|\cdot\|_{X}$ and $\|\cdot\|_{ Y}$, respectively, and $T: X\to Y$ is a compact operator. Let $\cN(\epsilon)$ denote the number of degrees of freedom of $T$ at level~$\epsilon$. Then
\begin{enumerate}
	\item \label{dofProp2} $\cN(\epsilon) = 0$ for all $\epsilon\geq\|T\|$.
	\item \label{dofProp5} Unless $T$ is identically zero, there
          exists an $\epsilon_0>0$ such that $\cN(\epsilon) \geq 1$
          for all $0<\epsilon<\epsilon_0$.      	
	\item \label{dofProp3} $\cN(\epsilon)$ is a non-increasing,
          upper semicontinuous function of $\epsilon$. 	
	\item \label{dofProp4} In any finite interval
          $(\epsilon_1,\epsilon_2)\subset\bbR$, with $0<\epsilon_1<
          \epsilon_2$, $\cN(\epsilon)$ has only finitely many
          discontinuities, i.e. $\cN(\epsilon)$ only takes finitely
          many non-negative integer values in any finite $\epsilon$ interval. 
\end{enumerate}
\end{theorem} 

The following two examples show that as $\epsilon$ goes to zero, $\cN(\epsilon)$ need not be finite nor go to infinity. 
\begin{example}\label{EDDBeg:dof1} Let $l^1$ be the Banach space of
  all real-valued sequences with finite $l^1$ norm and let
  $(e_1,e_2,\ldots)$ be the standard Schauder basis for $l^1$. Define
  the operator $T:l^1\to l^1$ by $e_n\mapsto e_1$ for all
  $n\in\bbZ^+$. This operator is well-defined and compact and
  $\cN(\epsilon)\leq 1$ for all $\epsilon > 0$. 
\end{example}
\begin{example} Let $l^1$ and $(e_1,e_2,\ldots)$ be defined as in the previous example. Define $T:l^1\to l^1$ by $e_n\mapsto \frac{1}{n} e_n$ for all $n\in \bbZ^+$. Again $T$ is well-defined and compact but $\lim_{\epsilon\to 0} \cN(\epsilon) = \infty$. 
\end{example}
\begin{figure}
\psfrag{ldot}{$\ddots$}\psfrag{epsilon}{$\epsilon$}
\includegraphics[width=0.5\textwidth]{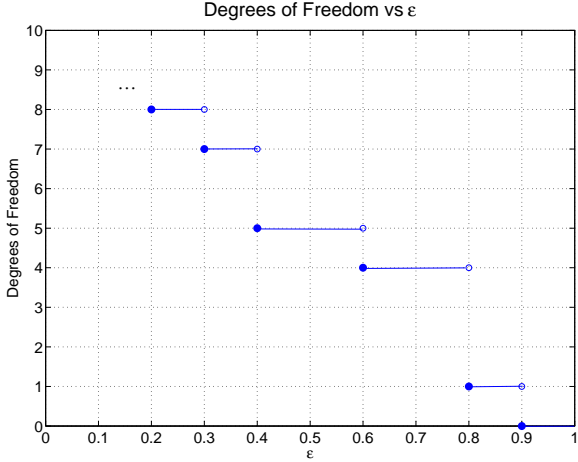}
\caption{Degrees of Freedom of a Compact Operator}\label{EDDBfig:dof}
\end{figure}
Figure~\ref{EDDBfig:dof} shows a typical example of degrees of freedom at level~$\epsilon$ for some compact operator that satisfies all the properties in the above theorem. 

\subsection{Generalised Singular Values} We will identify the
discontinuities in the number of degrees of freedom of $T$ at
level~$\epsilon$ with the (generalised) singular values of $T$.
\begin{definition}[Generalised Singular Values]\label{NCDdef:singVal} Suppose $ X$ and $ Y$ are normed spaces and $T: X\to Y$ is a compact operator. Let $\cN(\epsilon)$ denote the number of degrees of freedom of $T$ at level~$\epsilon$. Then $\epsilon_m$ is the $m^{th}$ generalised singular value of $T$ if
\begin{eqnarray*}
  &\sup_{\epsilon > \epsilon_m} \cN(\epsilon) = m-1& \textrm{and}\\
  &\inf_{\epsilon < \epsilon_m} \cN(\epsilon) = M \geq m.&
\end{eqnarray*}
Further, if $m<M$ then for all $m < n \leq M$, $\epsilon_n := \epsilon_m$ is the
$n^{th}$ generalised singular value of $T$.
\end{definition}
Note that by Theorem~\ref{EDDBthe:dofProp}, part~\ref{dofProp3} we
have $\cN(\epsilon_{m})\leq m-1$ with equality if (but not only if)
$\epsilon_{m}$ is not a repeated generalised singular value.

Let the degrees of freedom of some operator $T$ be as shown in
Figure~\ref{EDDBfig:dof}. Then the generalised singular values,
$\epsilon_m$, of $T$ identify the jumps in the degrees of freedom. So,
$\epsilon_1 = 0.9, \epsilon_2=\epsilon_3=\epsilon_4 = 0.8,\epsilon_5 =
0.6,\ldots$

Another way of understanding the connection between the number of
degrees of freedom at level~$\epsilon$ and generalised singular values
is as follows.
\begin{prop}\label{prop:numbers}
 Suppose $ X$ and $ Y$ are normed spaces and $T: X\to Y$ is a compact
 operator. Let $\cN(\epsilon)$ denote the number of degrees of freedom
 of $T$ at level~$\epsilon$. Then $\cN(\epsilon)$ is equal to the
 number of generalised singular values that are greater than $\epsilon$.  
\end{prop}
The intuition behind the definition for generalised singular values needs further clarification. In the finite dimensional case, if $\sigma_p$ is the $p^{th}$ singular value of some operator $T:\bbC^n\to\bbC^m$, then there exist corresponding left and right singular vectors $v_p\in \bbC^n$ and $u_p\in\bbC^m$ such that $v_p$ is of unit norm, $T v_p = u_p$ and the norm of $u_p$ is $\sigma_p$. This is not necessarily true for arbitrary compact operators on normed spaces as the following example proves. 
\begin{example}  Let $l^1$ and $(e_1,e_2,\ldots)$ be defined as in Example~\ref{EDDBeg:dof1}. Define the operator $T:l^1\to l^1$ by $T e_n = (1-\frac{1}{n}) e_1$ for all $n\in\bbZ^+$. Then $T$ is well-defined and compact. Also, the number of degrees of freedom of $T$ at level~$\epsilon$ is
\begin{equation*}
\cN(\epsilon) = \left\{\begin{array}{ll} 0 &\textrm{if } \epsilon \geq 1,\\ 1&\textrm{if } \epsilon < 1.\end{array}\right.
\end{equation*} 
So $\epsilon_1 = 1$. However, for any vector $x$ in the unit sphere in $l^1$, $\|T x\|_{l^1} < 1$. 
\end{example}
The above example motivates the slightly more complicated statement in
the following theorem which explains the intuition behind the definition of generalised singular values. 
\begin{theorem}\label{OMthe:new}
Suppose $ X$ and $ Y$ are normed spaces with norms $\|\cdot\|_{ X}$
and $\|\cdot\|_{ Y}$, respectively, and $T: X\to Y$ is a compact
operator. Let $\epsilon_m$ be a generalised singular value of the operator
$T$. Then for all $\theta>0$ there exists a $\phi\in X$, $\|\phi\|_{X}=1$, such that 
\begin{equation*}
\epsilon_m + \theta \geq \|T\phi\|_{ Y} \geq \epsilon_m-\theta.
\end{equation*}
\end{theorem}
The above theorem shows how the generalised singular values are
related to the traditionally accepted notion of singular values of
compact operators on Hilbert spaces.
In general, they are values the operator restricted to the unit sphere
can get arbitrarily close to in norm.
However, we still need to prove that in the special case of Hilbert
spaces the new definition for generalised singular values agrees with
the traditionally accepted definition for singular values.  

Recall that if $\cH_1$ and $\cH_2$ are Hilbert spaces with inner products $\iprod{\cdot}{\cdot}_{\cH_1}$ and $\iprod{\cdot}{\cdot}_{\cH_2}$ respectively and if $T:\cH_1\to \cH_2$ is a compact operator then the Hilbert adjoint operator for $T$ is defined as the unique operator $T^*:\cH_2\to \cH_1$ that satisfies~\cite[Sec. 3.9]{Kreyszig1989}
\begin{equation*}   
\iprod{Tx}{y}_{\cH_2} = \iprod{x}{T^*y}_{\cH_1}
\end{equation*}
for all $x\in \cH_1$ and $y\in\cH_2$.
The singular values of $T$ are defined to be the square roots of the eigenvalues of the operator $T^*T:\cH_1\to \cH_1$. We will refer to these as Hilbert space singular values to distinguish them from generalised singular values.  
Note that we always count repeated eigenvalues or (generalised) singular values repeatedly.
The following two theorems establish the connection between Hilbert
space singular values and the number of degrees of freedom at
level~$\epsilon$. The theorems are important in their own right
because they show that there are two other equivalent ways of
calculating the degrees of freedom of a Hilbert space operator.  
\begin{theorem}\label{Nthe:defHil} Suppose $\cH_1$ and $\cH_2$ are Hilbert spaces and
$T:\cH_1\to \cH_2$ is a compact operator. Then for all $\epsilon>
0$ there exist an $N\in \bbZ^+_0$ and a set of $N$ mutually
orthogonal vectors $\{\phi_i\}^N_{i=1}\subset\cH_{1}$ such that if
\begin{equation*}
    x \in \cH_1,\ \|x\|_{\cH_1}\leq 1\textrm{ and }\iprod{x}{\phi_i}_{\cH_1} = 0
\end{equation*}
then
\begin{equation*}
    \|Tx\|_{\cH_2}\leq \epsilon.
\end{equation*}
Moreover, the smallest $N$ that satisfies the above condition
for a given $\epsilon$ is equal to the number of Hilbert space
singular values of $T$ that are greater than $\epsilon$.
\end{theorem}
\begin{theorem}\label{Ncor:sinDof} Suppose that $\cH_1$ and $\cH_2$ are Hilbert
spaces and $T:\cH_1\to \cH_2$ is a compact operator. Then the
number of degrees of freedom at level~$\epsilon$ is equal to the
number of Hilbert space singular values of $T$ that are greater than $\epsilon$.
\end{theorem}
As a corollary of Theorem~\ref{Ncor:sinDof} we get the following result.
\begin{corollary}\label{EDDBthe:sVal} Suppose $\cH_1$ and $\cH_2$ are Hilbert spaces and
$T:\cH_1\to\cH_2$ is a compact operator. Suppose $\{\epsilon_m\}$
are the generalised singular values of $T$ and $\{\sigma_m\}$
are the possibly repeated Hilbert space singular values of $T$ written in non-increasing
order. Then
\begin{equation*}
    \sigma_m = \epsilon_m
\end{equation*}
for all $m\in\bbZ^{+}$.
\end{corollary}
This corollary, reassuringly, proves that the generalised singular values are in fact generalisations of the traditionally accepted notion of Hilbert space singular values. We will therefore use the terms generalised singular values and singular values interchangeably unless specified otherwise for the remainder of this paper. 

In Hilbert spaces we have three characterizations for degrees of freedom: 1) as in Definition~\ref{NCDdef:singVal}, 2) as in Theorem~\ref{Ncor:sinDof} in terms of singular values and 3) as in Theorem~\ref{Nthe:defHil} in terms of mutually orthogonal functions in the domain. 

We have used the first two characterisations in the generalisation to
normed spaces. However, the final characterisation is more difficult
to generalise. It  would be extremely useful to generalise the final
characterisation because, for the Hilbert space case, the functions
$\phi_i$ in Theorem~\ref{Nthe:defHil} are in some sense the best
functions to transmit (see e.g. ~\cite{Miller2000}). One could
possibly replace the mutual orthogonality by almost orthogonality using the Riesz lemma (see e.g. ~\cite[pp. 78]{Kreyszig1989}).
\begin{lemma}[Riesz's lemma] Let $Y$ and $Z$ be subspaces of a normed
  space $X$ and suppose that $Y$ is closed and is a proper subspace of
  $Z$. Then for all $\theta\in(0,1)$ there exists a $z\in Z$, $\|z\| = 1$, such that
  for all $y\in Y$
\begin{equation*}
	\|y-z\| \geq \theta.
\end{equation*}    
\end{lemma}
The following conjecture is still an open question.
\begin{conjecture}
Let $ X$ and $ Y$ be reflexive Banach spaces and let $T: X\to
 Y$ be compact. Given any $\epsilon> 0$ and some $\theta\in
(0,1)$, there exists a finite set of vectors
$\{\phi_i\}_{i=1}^N\subset X$
such that for all $x\in X$, $\|x\|_{ X} \leq 1$,
\begin{equation}\label{Neqn:almostOrth}
    \inf_{a_1,\ldots,a_N}
    \left\|x-\sum_{i=1}^N a_i\phi_i\right\|_{ X} \geq \theta
\end{equation}
implies
\begin{equation*}
    \|Tx\|_{ Y} \leq \epsilon.
\end{equation*}
\end{conjecture}
Comparing with Theorem~\ref{Nthe:defHil},
condition~\eqref{Neqn:almostOrth} is analogous to requiring that
$x$ be orthogonal to all the $\phi_i$. The conjecture is definitely not true unless we impose additional conditions such as reflexivity on $X$ and/or $Y$ as the next example proves.

\begin{example}
Let $l^1$, $(e_1,e_2,\ldots)$ and the compact operator $T:l^1\to l^1$ be defined as in Example~\ref{EDDBeg:dof1}. 
Now let $\epsilon <1$. For any $x = \sum_n \alpha_n e_n\in l^1$, if $\|x\|=1$ and
if $\alpha_{n}\geq 0$ for all $n$ then $\|Tx\| = \|x\| = 1 >\epsilon$. Hence no
finite set of vectors can satisfy the conditions in the conjecture.
\end{example}
In the following subsection, we use degrees of freedom and generalised singular values to define the essential dimension of a communication channel.
\subsection{Essential Dimension for Compact Operators}\label{EDdefDeg}
The definition for degrees of freedom given in Section~\ref{sec:degr-freed-comp} depends on the arbitrarily chosen number $\epsilon$ and therefore this definition does not give a unique number for a given channel. The physical intuition behind choosing this arbitrary small number $\epsilon$ is nicely explained in Xu and Janaswamy~\cite{Xu2006}. In that paper $\epsilon = \sigma^2$ denotes the noise level at the receiver and the authors state that the number of degrees of freedom fundamentally depends on this noise level.

However, in several important cases the number of degrees of freedom of a
channel is essentially independent of this arbitrarily chosen
positive
number~\cite{Migliore2006,Miller2000,Landau1962,Bucci1989,Poon2005,Kennedy2007}. This
is due to the fact that in these cases the singular values of the
channel operator show a step like behavior. Therefore, for a big range
of values of $\epsilon$, the number of degrees of freedom at
level~$\epsilon$ is constant. This leads us to the concept of
essential dimensionality\footnote{Note that the term ``essential
  dimension'' has been used instead of ``degrees of freedom'' in
  several papers. As far as we are aware, this is the first time an explicit
  distinction is being made between the two terms.} which is only a
function of the channel and not the arbitrarily chosen positive number
$\epsilon$. Some of the properties that one might require from the
essential dimension of a channel operator are:
\begin{enumerate}
    \item It must be uniquely defined for a given operator $T$.
    \item The definition must be applicable to a general class
    of operators under consideration so that comparisons can be made
    between different operators.\footnote{This requirement is in
      contrast to the essential dimension 
    definition in~\cite{Slepian1976} that is only applicable to the time-bandwidth problem.}
    \item It must in some sense \emph{represent} the number of degrees of
    freedom at level~$\epsilon$.
\end{enumerate}

The last requirement above needs
further clarification. Obviously the essential dimension of
$T$ can not in general be equal to the number of degrees of
freedom at level~$\epsilon$ because the latter is a function of
$\epsilon$. However, if the singular values of $T$ plotted in
non-increasing order change suddenly from being large to being small then
the number of degrees of freedom at the ``knee'' in this graph is the
essential dimension of $T$. The following 
definition for the essential dimension tries to identify this
``knee'' in the set of generalised singular values.

Each level~$\epsilon$ defines a unique number of degrees of
freedom $\cN(\epsilon)$ for a given compact operator
$T$. So for each positive integer $n\in\bbZ^+$ we can
calculate $E(n) = \mu(\{\epsilon: n = \cN(\epsilon)\})$. Here
$\mu(\cdot)$ is the Lebesgue measure. The function $E(n)$ is well
defined because of the properties of generalised singular values
discussed in Theorem~\ref{EDDBthe:dofProp}. We can now define the 
essential dimension of $T$ as follows.
\begin{definition}\label{EDDDdef:essDim} The essential dimension of a compact operator
$T$ is
\begin{equation*}
    \mathrm{EssDim}(T) = \mathrm{argmax}\{E(n): n \in
    \bbZ^+\}
\end{equation*}
where $E(n)$ is defined as above. If $\mathrm{argmax}$ above is
not unique then choose the smallest $n$ of all
the $n$ that maximise $E(n)$ as the essential dimension.
\end{definition}
In this definition we are simply calculating the maximum range of
values of the arbitrarily chosen $\epsilon$ over which the number
of degrees of freedom of an operator does not change.
It uniquely determines the essential dimension of all
compact operators. Further, it is equal to the number of degrees
of freedom at level~$\epsilon$ for the maximum range of
$\epsilon$. Choosing this value for the number of degrees of
freedom in order to model communication systems has the big advantage
that it is independent of the noise level at the receiver.
Further, if for a given noise level the number of degrees of
freedom is greater than the essential dimension then one can be
sure that even if the noise level varies by a significant amount
the number of degrees of freedom will always be greater than the
essential dimension.

The essential dimension of $T$ is the smallest number of generalised
singular values of $T$ after which the change in two consecutive
singular values is a maximum. One could also look at how the
generalised singular values are changing gradually and the above
definition is a special case of the following notion of essential
dimension of order $n$, namely the case where $n=1$.
\begin{definition}\label{EDDDdef:essDimN} Let $X,Y$ be normed spaces and let
$T:X\to Y$ be a compact operator. Let $\{\epsilon_m\}$ be the
set of generalised singular values of $T$ numbered in
non-increasing order. Then define the essential dimension of $T$
of order $n$ to be $N$ if $n$ is even and
\begin{equation*}
    \epsilon_{N-n/2}-\epsilon_{N+n/2} \geq \epsilon_{M-n/2}-\epsilon_{M+n/2}
\end{equation*}
for all $M\neq N$. If there are several $N$ that satisfy the above
condition then choose the smallest such $N$. If $n$ is odd then choose the smallest
$N$ that satisfies
\begin{equation*}
    \epsilon_{N-(n-1)/2}-\epsilon_{N+(n+1)/2} \geq \epsilon_{M-(n-1)/2}-\epsilon_{M+(n+1)/2}
\end{equation*}
for all $M\neq N$.
\end{definition}

A simple example illustrates the concepts of essential dimensionality and degrees of freedom.
\begin{example}
Figure~\ref{EDDNfig:sinVal} shows the singular values of some operator $T$. For this operator the number of degrees of freedom at level~$0.75$ is $7$ and at level~$0.1$ is $8$.  

The essential dimension of the channel is $7$. This is because for
$\epsilon \in [0.4,0.8)$, $\cN(\epsilon) = 7$. Therefore $E(7) = 0.4$
which is greater than $E(n)$ for all $n\ne 7$. The essential dimension
of order $2$ is $8$ because $\epsilon_7-\epsilon_9 = 0.7$ which is greater than $\epsilon_{M-1}-\epsilon_{M+1}$ for all $M\ne 8$. 
\begin{figure}
\begin{center}
  \includegraphics[width=0.5\textwidth]{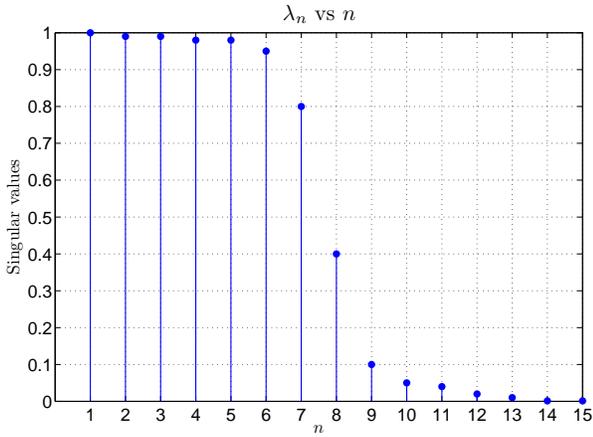}
  \caption{Singular values of an Operator}\label{EDDNfig:sinVal}
\end{center}
\end{figure}
\end{example}

\subsection{Computing generalised singular values}\label{NcomSin}
Both, degrees of freedom and essential dimension for a communication
channel, can be evaluated if the generalised singular values of the
operator $T$ describing the channel are known.  However, no known
method exists for computing these singular values for general compact
operators. In this section, we 
develop a numerical method, based on finite dimensional
approximations, that could be used to calculate generalised singular
values.  

\begin{theorem}\label{NCDthe:sinVal} Suppose $ X$ and $ Y$ are normed spaces
and $T: X\to Y$ is a compact operator. Also suppose that $ X$ has a
complete Schauder basis $\{\phi_1,\phi_2,\ldots\}$ and let
$S_n=\span\{\phi_1,\ldots,\phi_n\}$. Let $T_n = T|_{S_n}:S_n\to Y$,
$n \in \bbZ^+$. If $\epsilon_m$, the $m^{th}$ singular value of $T$, exists then for
$n$ large enough $\epsilon_{m,n}$, the $m^{th}$ singular value of
$T_n$, will exist and 
\begin{equation*}
    \lim_{n\to\infty}\epsilon_{m,n} = \epsilon_m.
\end{equation*}
If $\epsilon_{m,n}$ exists then it is a lower bound for
$\epsilon_m$.
\end{theorem}

The theorem shows that if the domain of the operator has some complete
Schauder basis then we can calculate the generalised singular values
of the operator restricted to finite dimensional subspaces and as the
subspaces get bigger we will approach the singular values of the
original operator. Moreover, the theorem also proves that the singular
values of the finite dimensional operators provide lower bounds for
the original generalised singular values. We, however, still need a practical method of calculating the singular values of linear operators defined on finite dimensional normed spaces.

Let $X,Y$ be two finite dimensional Banach spaces and let $T:X\to Y$
be a linear operator. Suppose $\epsilon_1,\ldots,\epsilon_n$ are the
generalised singular values of $T$ and denote $B_1 = \{x\in X:\|x\|_X \leq 1\}$.
We know that for all $\epsilon\geq\epsilon_{p+1}$, $\cN(\epsilon)\leq p$. Hence for each
$\epsilon\geq\epsilon_{p+1}$ there exists a set $\{\psi_i\}_{i=1}^p \subset Y$ such that  
\begin{equation*}
\sup_{x\in B_1}\inf_{a_1,\ldots,a_{p}} \left\|Tx - \sum_{i=1}^p a_i\psi_i\right\|_Y \leq \epsilon.
\end{equation*}
Let $\Psi_{p,\epsilon}$ denote the set of all sets $\{\psi_i:
\|\psi_i\|_Y \leq 1\}_{i=1}^p\subset Y$ that satisfy the above inequality for a given $\epsilon\geq\epsilon_{p+1}$ and let 
\begin{equation*}
\Psi_p = \bigcup_{\epsilon\geq\epsilon_{p+1}} \Psi_{p,\epsilon}.
\end{equation*}
With this notation we can now prove that the generalised singular
values of a linear operator defined on a finite dimensional normed
space can be expressed as the solution of an optimisation problem. 
\begin{theorem}\label{CScomSinTh:calc}
Let $X,Y$ be two finite dimensional Banach spaces and let $T:X\to Y$ be a linear operator. Also let $B_1$ be the closed unit ball in $X$ and suppose $\Psi_p$ is defined as explained above. Then
\begin{equation*} 
\sup_{x\in B_1} \|Tx\|_Y = \epsilon_1
\end{equation*}
and for all $p\in\bbZ^{+}$
\begin{equation*}
\inf_{\{\psi_i\}_{i=1}^p\in\Psi_p}\sup_{x\in B_1}\inf_{a_1,\ldots,a_{p}} \left\|Tx-\sum_{i=1}^p a_i\psi_i\right\|_{Y} = \epsilon_{p+1}.
\end{equation*}
\end{theorem}
Given the ``correct'' set of functions $\psi_{i}$, the above theorem
characterises the singular values in terms of a maximisation problem
over a finite dimensional domain. It is however difficult to check whether a given set of functions $\{\psi_i\}_{i=1}^p$ is an element of $\Psi_p$. We therefore propose the following algorithm to calculate bounds on the generalised singular values. 

Suppose $X,Y$, $T:X\to Y$, $\epsilon_1,\ldots,\epsilon_n$ and $B_1$
are defined as in Theorem~\ref{CScomSinTh:calc}. Let 
\begin{equation*}
\epsilon'_1 = \sup_{x\in B_1} \|Tx\|_Y.
\end{equation*}
Because $B_1\subset X$ is a compact set and $\|\cdot\|_Y$ and $T$ are continuous, there exists an $x_1\in B_1$ such that $\|Tx_1\|_Y = \epsilon_1'$. Choose $\psi_1 = Tx_1$. 

Now suppose $\psi_1,\ldots,\psi_p$ have been chosen. Then let 
\begin{equation}\label{CScomSinEq:best}
\epsilon'_{p+1}=\sup_{x\in B_1}\inf_{a_1,\ldots,a_{p}} \left\|Tx-\sum_{i=1}^p a_i\psi_i\right\|_{Y}.
\end{equation}
Again, because $B_1\subset X$ is a compact set and $\|\cdot\|_Y$ and $T$ are continuous, there exists an $x_{p+1}\in B_1$ such that $x_{p+1}$ attains the maximum in the above equation. Choose $\psi_{p+1} = T x_{p+1}$. Comparing with Theorem~\ref{CScomSinTh:calc} we note that $\epsilon'_{p+1}$ is an upper bound for $\epsilon_{p+1}$. It is an open question as to whether $\epsilon'_{p+1} = \epsilon_{p+1}$. 

In this algorithm, instead of searching over all possible sets in
$\Psi_p$ we select a special set that is in some sense (it consists of
images of the $x\in B_1$ that attain the maximum in equation~\eqref{CScomSinEq:best}) the best possible set to use. This choice is essential because otherwise the calculation of generalised singular values becomes too cumbersome (one needs to find the set $\Psi_p$ before calculating $\epsilon_{p+1}$). Note however, that the above algorithm gives the correct value for $\epsilon_1$. 

The theory presented here has been used to compute the generalised
singular values and degrees of freedom  in spatial waveform channels
of the type discussed in Example~\ref{eg:swc}. The results of these
computations are presented in Somaraju~\cite{thesis}. Due to space constraints, these results
are not further discussed in this paper.  

\subsection{Non-compactness of channel operators} Throughout this
paper we have exclusively dealt with channels that can be modeled
using compact operators. We have done so because of the
following result.
\begin{theorem}\label{Nthe:comp} (Converse to Theorem~\ref{Nthe:defBan})
Suppose $X$ and $Y$ are normed spaces with norms $\|\cdot\|_{ X}$ and
$\|\cdot\|_{ Y}$, respectively, and $T: X\to  Y$ is a bounded linear
operator. If for all $\epsilon > 0$ there exist $N\in \bbZ_0^+$ and
a set $\{\psi_i\}_{i=1}^N\subset  Y$ such that for all $x\in \bar{B}_{1,X}(0)$
\begin{equation*}
    \inf_{a_1,\ldots,a_N} \left\|Tx-\sum_{i=1}^N a_i\psi_i\right\|_{ Y}
    \leq \epsilon
\end{equation*}
then $T$ is compact.
\end{theorem}
So any bounded channel operator with finitely many sub-channels must
be compact. Indeed, if one can 
find a channel that is not described by a compact operator, then it
will have infinitely many sub-channels and will therefore have
infinite capacity. Also, if the channel is described by an operator
that is linear but unbounded then there will obviously exist
sub-channels over which arbitrarily large gains can be obtained.\footnote{It could hence be
  argued that non-compact channel operators are unphysical, however,
  we will leave it to the reader to make this judgement.} 

%%% Local Variables: 
%%% mode: latex
%%% TeX-master: "main"
%%% End: 

\section{Conclusion}\label{chaCon} In this paper we assume that a
communication channel can be modeled by a normed space $X$ of
transmitter functions that a transmitter can generate, a normed space
$Y$ of functions that a receiver can measure and an operator $T:X\to
Y$ that maps the transmitter functions to functions measured by the
receiver. We then introduce the concepts of degrees of freedom at
level~$\epsilon$, essential dimension and generalised singular values
of such channel operators in the case where they are compact. One can
give a physical interpretation for 
degrees of freedom as follows: if there is some constraint
$\|\cdot\|_X\leq 1$ on the space of source functions and if the
receiver can only measure signals that satisfy $\|\cdot\|_Y >
\epsilon$ then the number of
degrees of freedom is the number of linearly independent signals that
the receiver can measure under the given constraints. If the degrees of freedom are largely
independent of the level~$\epsilon$ then it makes sense to talk about
the essential dimension of the channel. The essential dimension of the
channel is the smallest number of degrees of freedom of the channel
that is the same for the
largest range of levels~$\epsilon$. We show how one can use the number
of degrees of freedom at level~$\epsilon$ to generalise the Hilbert
space concept of singular values to arbitrary normed spaces. We also
provide a simple algorithm that can be used to 
approximately calculate these generalised singular values.
Finally, we prove that if the operator describing the
channel is not compact then it must either have infinite gain or have
an infinite number of degrees of freedom. The general theory developed in this paper is
applied to spatial waveform channels in Somaraju~\cite{thesis}.  

%%% Local Variables: 
%%% mode: latex
%%% TeX-master: "main"
%%% End: 

\appendix{Proofs of Theorems:}
\emph{Theorem~\ref{Nthe:defBan}.} Suppose $X$ and $ Y$ are normed
spaces with norms $\|\cdot\|_{ X}$ and $\|\cdot\|_{ Y}$, respectively,
and $T: X\to  Y$ is a compact operator. Then for all $\epsilon > 0$
there exist $N\in \bbZ_0^+$ and a set $\{\psi_i\}_{i=1}^N\subset  Y$
such that for all $x\in \bar{B}_{1,X}(0)$
\begin{equation}\label{Neqn:defBan1}
    \inf_{a_1,\ldots,a_N} \left\|Tx-\sum_{i=1}^N a_i\psi_i\right\|_{ Y}
    \leq \epsilon.
\end{equation}
\begin{proof}
The proof is by contradiction. Let $\epsilon> 0$ be given. Suppose
no such $N$ exists.

Let $x_1\in  \bar{B}_{1,X}(0)$ be any vector. Choose $\psi_1 = Tx_1$.
Suppose that $\{x_1,\ldots,x_N\}$ and $\{\psi_1,\ldots,\psi_N\}$ 
have been chosen. Then, by our assumption, there exists an $x_{N+1}\in  \bar{B}_{1,X}(0)$ such that 
\begin{equation}\label{Neqn:defBan2}
    \inf_{a_1,\ldots,a_N} \left\|Tx_{N+1}-\sum_{i=1}^N a_i\psi_i\right\|_{ Y}
    > \epsilon.
\end{equation}
Choose $\psi_{N+1} = Tx_{N+1}$. By induction, for $M\leq N$ we have
\begin{equation*}
    \|Tx_{N+1}-Tx_M\|_{ Y} > \epsilon.
\end{equation*}
This follows from~\eqref{Neqn:defBan2} by setting $a_{i}=0$, $i\leq
N$, $i\ne M$, and $a_{M}=1$.
Therefore, using the Cauchy criterion, the sequence $\{Tx_n\}_{n=1}^\infty$ chosen by
induction cannot have a convergent subsequence. This is the
required contradiction because $\{x_n\}_{n=1}^\infty$ is a bounded
sequence and $T$ is compact.
\end{proof}

\emph{Theorem~\ref{EDDBthe:dofProp}.}
Suppose $ X$ and $ Y$ are normed spaces with norms $\|\cdot\|_{X}$ and $\|\cdot\|_{ Y}$, respectively, and $T: X\to Y$ is a compact operator. Let $\cN(\epsilon)$ denote the number of degrees of freedom of $T$ at level~$\epsilon$. Then
\begin{enumerate}
	\item $\cN(\epsilon) = 0$ for all $\epsilon\geq\|T\|$. 
	\item Unless $T$ is identically zero, there
          exists an $\epsilon_0>0$ such that $\cN(\epsilon) \geq 1$
          for all $0<\epsilon<\epsilon_0$.      	
	\item $\cN(\epsilon)$ is a non-increasing,
          upper semicontinuous function of $\epsilon$. 	
	\item In any finite interval
          $(\epsilon_1,\epsilon_2)\subset\bbR$, with $0<\epsilon_1<
          \epsilon_2$, $\cN(\epsilon)$ has only finitely many
          discontinuities, i.e. $\cN(\epsilon)$ only takes finitely
          many non-negative integer values in any finite $\epsilon$ interval. 
\end{enumerate}
\begin{proof}
\begin{enumerate}
	\item Because $T$ is compact it is bounded, and
          therefore $\|T\|<\infty$. Suppose $\epsilon\geq\|T\|$ then
          $\|Tx\|_{Y}\leq\|T\|\leq\epsilon$ for all $x\in
          \bar{B}_{1,X}(0)$. Therefore $\cN(\epsilon) = 0$.   
	\item If $\|T\|> 0$ there exists an $x\in X$, $\|x\|_{X}\le
          1$ such that $\|Tx\|_Y > 0$. Set
          $\epsilon_{0}:=\|Tx\|_Y$. Then for all
          $0<\epsilon<\epsilon_{0}$, $\cN(\epsilon)\geq 1$.  
	\item Suppose $0<\epsilon_1<\epsilon_2$. Then there exist
          functions $\psi_1,\ldots,\psi_{\cN(\epsilon_1)}$ such that
          for all $x\in  \bar{B}_{1,X}(0)$ 
	\begin{equation*}
		\inf_{a_1,\ldots,a_{\cN(\epsilon_1)}}
                \left\|Tx-\sum_{i=1}^{\cN(\epsilon_1)}a_i\psi_i\right\|_{Y}
                <\epsilon_1 <\epsilon_2 
	\end{equation*}
	Therefore $\cN(\epsilon_2)\leq\cN(\epsilon_1)$ from the
        definition of the number of degrees of freedom at
        level~$\epsilon$, i.e. $\cN(\epsilon)$ is non-increasing. In
        particular we have
        \begin{equation*}
          \lim_{\epsilon\searrow\epsilon_{1}}\cN(\epsilon)\leq\cN(\epsilon_{1}).
        \end{equation*}
        Assume that the above inequality is strict. Then there exists
        an $N\in\bbZ_{0}^{+}$, $N<\cN(\epsilon_{1})$, and for all $\theta>0$ there
        exists a set $\{\psi_{i}^{\theta}\}_{i=1}^{N}\subset Y$ such
        that for all $x\in\bar{B}_{1,X}(0)$
        \begin{equation}\label{eq:1}
          \inf_{a_1,\ldots,a_N} \left\|Tx-\sum_{i=1}^N a_i\psi_i^{\theta}\right\|_{ Y}
          \leq \epsilon_{1}+\theta.
        \end{equation}
        On the other hand, since $\cN(\epsilon_{1})>N$, for all sets
        $\{\psi_{i}\}_{i=1}^{N}\subset Y$ there exists an
        $x\in\bar{B}_{1,X}(0)$ such that
        \begin{equation}
          \label{eq:2}
           \mu:=\inf_{a_1,\ldots,a_N} \left\|Tx-\sum_{i=1}^N a_i\psi_i\right\|_{ Y}
           >\epsilon_{1}.
        \end{equation}
        But~\eqref{eq:1} contradicts~\eqref{eq:2} for
        $\theta:=\frac{1}{2}(\mu-\epsilon_{1})$. Hence
        $\lim_{\epsilon\searrow\epsilon_{1}}\cN(\epsilon)=\cN(\epsilon_{1})$
        and $\cN(\epsilon)$ is upper semicontinuous.
	\item This follows from Parts~\ref{dofProp2} and~\ref{dofProp3}.
\end{enumerate}
\end{proof}

\emph{Proposition~\ref{prop:numbers}.}
 Suppose $ X$ and $ Y$ are normed spaces and $T: X\to Y$ is a compact
 operator. Let $\cN(\epsilon)$ denote the number of degrees of freedom
 of $T$ at level~$\epsilon$. Then $\cN(\epsilon)$ is equal to the
 number of generalised singular values that are greater than $\epsilon$.  

 \begin{proof}
   This follows from careful counting of the numbers of degrees of
   freedom at level~$\epsilon$ including repeated counting according
   to the height of any occurring ``jumps''. 
 \end{proof}

\emph{Theorem~\ref{OMthe:new}.}
Suppose $ X$ and $ Y$ are normed spaces with norms $\|\cdot\|_{ X}$
and $\|\cdot\|_{ Y}$, respectively, and $T: X\to Y$ is a compact
operator. Let $\epsilon_m$ be a generalised singular value of the operator
$T$. Then for all $\theta>0$ there exists a $\phi\in X$, $\|\phi\|_{X}=1$, such that 
\begin{equation*}
\epsilon_m + \theta \geq \|T\phi\|_{ Y} \geq \epsilon_m-\theta.
\end{equation*}
\begin{proof} The proof is by contradiction. Assume that there exists
  a $\theta > 0$ such that for all $\phi\in  X$, $\|\phi\|_{ X} = 1$,
  we have $\|T \phi\|_{ Y} \notin
  [\epsilon_m-\theta,\epsilon_m+\theta]$. Let $\cN(\epsilon)$ denote
  the number of degrees of freedom at level~$\epsilon$ of the operator
  $T$.  From the definition of degrees of freedom at level~$\epsilon$
  we have 
\begin{eqnarray}
\cN(\epsilon_m + \theta) &\leq& m-1,\label{eq2}\\
\cN(\epsilon_m - \theta) &\geq& m.\label{eq1}
\end{eqnarray}
By~\eqref{eq2}, there exist vectors $\psi_1,\ldots,\psi_{m-1}\in Y$ such that for all $x\in \bar{B}_{1,X}(0)$
\begin{equation*}
\inf_{a_1,\ldots,a_{m-1}} \left\|Tx - \sum_{i=1}^{m-1} a_i\psi_i\right\| \leq \epsilon_m + \theta.
\end{equation*}
By our assumption on $\|T \phi\|_{ Y}$,
\begin{equation*}
\inf_{a_1,\ldots,a_{m-1}} \left\|T\phi - \sum_{i=1}^{m-1} a_i\psi_i\right\| \leq \epsilon_m - \theta.
\end{equation*}
This follows from consideration of the case
$a_{1}=\dots=a_{m-1}=0$. Hence $\cN(\epsilon_m-\theta)\leq m-1$ since
scaling $\phi$ to non-unit norm is equivalent to scaling all the
$a_{i}$. This contradicts inequality~\eqref{eq1}. Therefore there
exists a $\phi$ that satisfies the conditions of the theorem.  
\end{proof}

\emph{Theorem~\ref{Nthe:defHil}.} Suppose $\cH_1$ and $\cH_2$ are Hilbert spaces and
$T:\cH_1\to \cH_2$ is a compact operator. Then for all $\epsilon>
0$ there exist an $N\in \bbZ^+_0$ and a set of $N$ mutually
orthogonal vectors $\{\phi_i\}^N_{i=1}\subset\cH_{1}$ such that if
\begin{equation*}
    x \in \cH_1,\ \|x\|_{\cH_1}\leq 1\textrm{ and }\iprod{x}{\phi_i}_{\cH_1} = 0
\end{equation*}
then
\begin{equation*}
    \|Tx\|_{\cH_2}\leq \epsilon.
\end{equation*}
Moreover, the smallest $N$ that satisfies the above condition
for a given $\epsilon$ is equal to the number of Hilbert space
singular values of $T$ that are greater than $\epsilon$.

\begin{proof} We first prove that such an $N$ is given by the number of
Hilbert space singular values of $T$ that are greater than $\epsilon$
and then prove that this is the smallest such $N$.

Let $\epsilon>0$ be given. Because $T$ is compact, we can use the
singular value decomposition theorem which says \cite[p.~261]{Kato1980}
\begin{equation}\label{Neqn:svd}
    T\cdot = \sum_i \sigma_i \iprod{\cdot}{\phi_i}_{\cH_1}\psi_i.
\end{equation}
Here, $\sigma_i$, $\phi_i$ and $\psi_i$ with $i\in\bbZ^{+}$ are the
Hilbert space singular values and 
left and right singular vectors of $T$, respectively. We assume
w.l.o.g. that the Hilbert space singular values are ordered in
non-increasing order. We denote by $N_1\in\bbZ^+$ the
number of Hilbert space singular values of $T$ that are greater than
$\epsilon$, i.e. $\sigma_{i}>\epsilon$ if and only if $i\leq N_{1}$.

Now, if $x$ is orthogonal to $\phi_i,i =1,\ldots,N_1$ and if
$\|x\|_{\cH_1}\leq 1$ then from equation~\eqref{Neqn:svd}
\begin{eqnarray*}
  \|Tx\|_{\cH_2}^2 &=& \sum_{i=1}^\infty\sigma_i^2|\iprod{x}{\phi_i}_{\cH_1}|^2 \|\psi_i\|_{\cH_2}^2 \\
   &\leq& \epsilon^2\sum_{i=N_1+1}^\infty|\iprod{x}{\phi_i}_{\cH_1}|^2 \\
   &\leq& \epsilon^2.
\end{eqnarray*}
For $N<N_{1}$, the linear span of any set
$\{\varphi_{i}\}_{i=1}^{N}\subset\cH_{1}$ has a non-trivial orthogonal
complement
in the span of $\{\phi_{i}\}_{i=1}^{N_{1}}$. Any vector $x$ in this
complement with $\|x\|_{\cH_{1}}=1$ fullfills the conditions of
the theorem but $\|Tx\|_{\cH_{2}}>\epsilon$ by equation~\eqref{Neqn:svd}.
\end{proof}

\emph{Theorem~\ref{Ncor:sinDof}.} Suppose that $\cH_1$ and $\cH_2$ are Hilbert
spaces and $T:\cH_1\to \cH_2$ is a compact operator. Then the
number of degrees of freedom at level~$\epsilon$ is equal to the
number of Hilbert space singular values of $T$ that are greater than
$\epsilon$.

\begin{proof}
As in the prove of the previous theorem, let $N_{1}\in\bbZ^{+}$ denote
the number of Hilbert space singular 
values of $T$ that are greater than $\epsilon$. Let $\sigma_i$,
$\phi_i$ and $\psi_i$ with $i\in\bbZ^{+}$ denote the Hilbert space
singular values in non-increasing order and the
left and right singular vectors of $T$, respectively. Let
$N_{2}\in\bbZ^{+}$ denote the
number of degrees of freedom of $T$ at level~$\epsilon$. 

We first prove that $N_1\geq N_2$. If $x$ is in the unit ball in
${\cH_1}$ then we can write $x=\sum_{i=1}^\infty
\iprod{x}{\phi_i}_{\cH_1}\phi_i + x_r$. Here $x_r$ is the
remainder term that is orthogonal to all the $\phi_i$. From
equation~\eqref{Neqn:svd} and $\sigma_{i}\leq\epsilon$ for $i>N_{1}$
it follows that
\begin{equation*}
    \left\|Tx-\sum_{i=1}^{N_1}\sigma_i\iprod{x}{\phi_i}_{\cH_1} \psi_i\right\|_{\cH_2}
    \leq \epsilon
\end{equation*}
and hence $N_{1}\geq N_{2}$ by the definition of the number of degrees
of freedom at level $\epsilon$ (set
$a_{i}=\sigma_i\iprod{x}{\phi_i}_{\cH_1}$ in that definition).

To prove that $N_1\leq N_2$ assume that $N_1>N_2$ to arrive at a
contradiction. Then there exists a set $\{\psi_i'\}_{i=1}^{N_2}\subset\cH_{2}$
such that
\begin{equation*}
    \inf_{a_1,\ldots,a_{N_{2}}} \left\|Tx-\sum_{i=1}^{N_2} a_i\psi'_i\right\|_{\cH_2}
    \leq \epsilon
\end{equation*}
for all $x\in\cH_1$, $\|x\|_{ \cH_1}\leq 1$.
Because we assume that $N_1>N_2$, there exists a $y\in
\mathrm{span}\{\psi_1,\ldots,\psi_{N_1}\}$ which is orthogonal to
all the $\psi_i'$. Let $y = \sum_{i=1}^{N_1} b_i\psi_i$. Then $y = Tx$
where $x = \sum_{i=1}^{N_1} \frac{b_i}{\sigma_i}\phi_i$ by
equation~\eqref{Neqn:svd}. We can assume w.l.o.g. that the $b_i$ 
are normalised so that $\|x\|_{\cH_1} = 1$. If this is done then
\begin{eqnarray}
   \inf_{a_1,\ldots,a_{N_{2}}} \left\|Tx-\sum_{i=1}^{N_2}
   a_i\psi'_i\right\|^2_{ \cH_2} &=& \|y\|_{\cH_2}^2\label{EDDBeq:step1}\\
   &=& \sum_{i=1}^{N_1} b_i^2 \label{EDDBeq:step2}\\
    &>& \sum_{i=1}^{N_1} \frac{b_i^2}{\sigma_i^2}\epsilon^2\label{EDDBeq:step3}\\
    &=& \epsilon^2\label{EDDBeq:step4}.
\end{eqnarray}
In the above we get equation~\eqref{EDDBeq:step1} from the fact that $y$
is orthogonal to all the $\psi_i'$, inequality~\eqref{EDDBeq:step3}
from $\sigma_{i}>\epsilon$ for $i\leq N_{1}$ and
equation~\eqref{EDDBeq:step4} from $\|x\|_{\cH_1} = 1$.
The inequality~\eqref{EDDBeq:step1}--\eqref{EDDBeq:step4} is the required contradiction. This
proves that $N_{1}\leq N_{2}$ and hence $N_1 = N_2$.
\end{proof}

\emph{Corollary~\ref{EDDBthe:sVal}.} Suppose $\cH_1$ and $\cH_2$ are Hilbert spaces and
$T:\cH_1\to\cH_2$ is a compact operator. Suppose $\{\epsilon_m\}$
are the generalised singular values of $T$ and $\{\sigma_m\}$
are the possibly repeated Hilbert space singular values of $T$ written in non-increasing
order. Then
\begin{equation*}
    \sigma_m = \epsilon_m
\end{equation*}
for all $m\in\bbZ^{+}$.

\begin{proof} 
This follows immediately from Theorem~\ref{Ncor:sinDof} and
Proposition~\ref{prop:numbers} by a simple counting argument.
\end{proof}

\emph{Theorem~\ref{NCDthe:sinVal}.}  Suppose $ X$ and $ Y$ are normed spaces
and $T: X\to Y$ is a compact operator. Also suppose that $ X$ has a
complete Schauder basis $\{\phi_1,\phi_2,\ldots\}$ and let
$S_n=\span\{\phi_1,\ldots,\phi_n\}$. Let $T_n = T|_{S_n}:S_n\to Y$,
$n \in \bbZ^+$. If $\epsilon_m$, the $m^{th}$ singular value of $T$, exists then for
$n$ large enough $\epsilon_{m,n}$, the $m^{th}$ singular value of
$T_n$, will exist and 
\begin{equation*}
    \lim_{n\to\infty}\epsilon_{m,n} = \epsilon_m.
\end{equation*}
If $\epsilon_{m,n}$ exists then it is a lower bound for
$\epsilon_m$.\newline
\emph{Proof Outline:} The crux of the argument used to prove the
theorem is as follows.  Assume $\epsilon>0$ is given and let
$\cN(\epsilon)$ denote the number of degrees of freedom at
level~$\epsilon$ for the operator $T$. By definition there exist
functions $\{\psi_1,\ldots,\psi_{\cN(\epsilon)}\}\subset Y$ such that
for all $x\in X$, $\|x\|_{X}\leq 1$, $Tx$ can be approximated to level~$\epsilon$
by a linear combination of the $\psi_i$ and further, no set of
functions $\{\psi'_1,\ldots,\psi'_N\} \subset Y$ can approximate all the
$Tx$ if $N<\cN(\epsilon)$. Equivalently, there is a vector
in the closed unit ball in $X$ whose image under $T$ can be approximated by a
vector in $\span\{\psi_1,\ldots,\psi_{\cN(\epsilon)}\}$ but not by any
vector in $\span\{\psi'_1,\ldots,\psi'_{N}\}$.   

So we take the inverse image of an $\epsilon$-net of points in
$\span\{\psi_1,\ldots,\psi_{\cN(\epsilon)}\}$ and choose $n$ large
enough so that all the inverse images are close to $S_n$. We can do
this because the $\phi_i$ form a complete Schauder basis for $X$. We
then show that there exists a vector in $S_n$ such that its image
under $T$ cannot be approximated by a linear combination of
$\psi'_1,\ldots,\psi'_N$ for $N < \cN(\epsilon)$. This will prove that
the number of degrees of freedom at level~$\epsilon$ of $T_n$
approaches that of $T$ and consequently so do the singular values. The
details are as follows.  

\begin{proof} 
We will prove this theorem in two parts. Assume that $\epsilon_{m}$
exists. In part a) we
will prove that if $\epsilon_{m,N}$ exists for some $N\in\bbZ^{+}$
then $\epsilon_{m,n}$ exists  for all $n>N$, and the $\epsilon_{m,n}$
form a non-decreasing sequence indexed by $n$ that is bounded from above by
$\epsilon_m$. In part b) we prove by contradiction that
$\epsilon_{m,n}$ exists for some $n\in\bbZ^{+}$ and that $\epsilon_{m,n}$
must converge to $\epsilon_m$. 

We will use the following notation in the proof:
\begin{multline*}
\spe\{\psi_1,\ldots,\psi_N\} = \\ \{y\in Y: \inf_{a_1,\ldots,a_N}
\left\|y-\sum_{i=1}^N a_i\psi_i\right\|_{Y} \leq \epsilon\}
\end{multline*}
and $B_r = \{x\in X:\|x\|_{ X}\leq r\}$.

\textbf{Part a:} Let $T$ and $T_n$ be defined as in the theorem and let $\cN(\epsilon)$ and $\cN_n(\epsilon)$ be the numbers of degrees of freedom at level~$\epsilon$ of $T$ and $T_n$, respectively. Assume that $\epsilon_{m,n_1}$ exists and let $n_2>n_1$. 

Then for all sets $\{\psi_1,\ldots,\psi_{\cN_{n_1}(\epsilon)-1}\}\subset Y$ there is a $\xi\in S_{n_1}\cap B_1$ such that
\begin{equation*}
	T_{n_{1}}\xi = T\xi \notin \spe\{\psi_1,\ldots,\psi_{\cN_{n_1}(\epsilon)-1}\}.
\end{equation*}
Because $S_{n_1}\subset S_{n_2}$ we have $\xi\in S_{n_2} \cap B_1$  and\begin{equation*}
	T_{n_2}\xi = T\xi \notin \spe\{\psi_1,\ldots,\psi_{\cN_{n_1}(\epsilon)-1}\}.
\end{equation*}
Therefore for all $\epsilon>0$
\begin{equation}\label{NCDeqn:cN}
	\cN_{n_2}(\epsilon) \geq \cN_{n_1}(\epsilon).
\end{equation}
Because
\begin{equation}\label{SCCSeq:joc}
\inf_{\epsilon < \epsilon_{m,n_1}} \cN_{n_1}(\epsilon) \geq m 
\end{equation}
we have $\cN_{n_2}(\epsilon) \ge \cN_{n_1}(\epsilon) \geq m$ for $\epsilon<\epsilon_{m,n_1}$. Hence $\epsilon_{m,n_2}$ must exist.

From the definition of generalised singular values we have inequality~\eqref{SCCSeq:joc} and 
\begin{eqnarray*}
	&\sup_{\epsilon > \epsilon_{m,n_2}} \cN_{n_2}(\epsilon) \leq m-1&
\end{eqnarray*}
If $\epsilon_{m,n_1} > \epsilon_{m,n_2}$ then there exists an $\epsilon'$ such that $\epsilon_{m,n_1} > \epsilon' > \epsilon_{m,n_2}$. Therefore,
\begin{equation*}
	\cN_{n_1}(\epsilon') \geq m > m-1\geq \cN_{n_2}(\epsilon').
\end{equation*}
This contradicts inequality~\eqref{NCDeqn:cN}. Therefore $\epsilon_{m,n_1} \leq \epsilon_{m,n_2}$. 

The same line of arguments as above can be used to show that if both
$\epsilon_{m}$ and $\epsilon_{m,n}$ exist then
$\epsilon_{m,n}\leq\epsilon_{m}$.
Recall that we have assumed at the beginning that $\epsilon_{m}$ exists.
Therefore, if $\epsilon_{m,N}$ exists for some $N\in \bbZ^{+}$ then $\epsilon_{m,n}$ is a non-decreasing sequence in $n\geq N$ that is bounded from above by $\epsilon_m$.\newline 
\textbf{Part b:}  
By part a), if $\epsilon_{m,n}$ exists for $n\geq n_1$ then, because $\epsilon_{m,n}$ is a bounded monotonic sequence in $n$ it must converge to some $\epsilon_m'\leq \epsilon_m$.

Now there are two situations to consider. Firstly, $\epsilon_{m,n}$
might not exist for any $n\in \bbZ^{+}$. Secondly, $\epsilon_{m,n}$ might
exist for some $n$ but the limit $\epsilon_m'$ might be strictly less
than $\epsilon_m$. We consider the two situations separately and
arrive at the same set of inequalities in both situations. We then derive a contradiction
from that set. 

\textbf{Situation 1:} Assume that $\epsilon_{m,n}$ does not exist for any $n\in \bbZ^{+}$. Then
\begin{equation}\label{e8}
\cN_n(\epsilon)\leq m-1
\end{equation}
for all $n\in \bbZ^{+}$ and $\epsilon>0$.
Using the definition of degrees of freedom for $T$ there exist constants
$\alpha <\beta <\epsilon_m$ such that  
\setlength{\arraycolsep}{0pt}
\begin{eqnarray}	
  &\cN_n(\alpha)\leq m-1& \quad \textrm{for all } n\in\bbZ^{+} \quad \textrm{and} \label{ea5}\\
  &\cN(\beta) \geq m.&\label{ea4}
\end{eqnarray}
\setlength{\arraycolsep}{5pt}

\textbf{Situation 2:} Assume that $\epsilon_m'<\epsilon_m$. From the definition of generalised singular values we know
\begin{eqnarray*}
  &\sup_{\epsilon > \epsilon_{m,n}} \cN_{n}(\epsilon) \leq m-1& \
  \textrm{for all } n\in\bbZ^{+} \quad \textrm{and}\\
  &\inf_{\epsilon < \epsilon_{m}} \cN(\epsilon) \geq m.&
\end{eqnarray*} 
Because $\epsilon_{m,n}\leq \epsilon'_m$, we know that there exist numbers $\alpha$ and $\beta$, $\epsilon'_m<\alpha<\beta<\epsilon_m$ such that
\begin{eqnarray}	
	&\cN_n(\alpha)\leq m-1& \ \textrm{for all } n\in\bbZ^{+} \quad \textrm{and}\label{e5}\\
	&\cN(\beta) \ge m.&\label{e4}
\end{eqnarray}
These are the same conditions as~\eqref{ea5} and~\eqref{ea4}.
Therefore, in both situations we need to prove that the inequalities~\eqref{e5} and~\eqref{e4} cannot be simultaneously true. 

Because $T$ is compact, $TB_1$ is totally
bounded~\cite[ch. 8]{Kreyszig1989}. Therefore, $TB_1$ has a finite
$\epsilon$-net for all $\epsilon>0$. Hence there exists a set of vectors $\{\xi_1,\ldots,\xi_P\}\subset B_1$ such that for all $y\in TB_1$ there exists a $p$, $1\leq p\leq P$ with
\begin{equation}\label{e1}	
	\|T\xi_p-y\|_{ Y} < \frac{\beta-\alpha}{2}. 
\end{equation} 
Now, because $\{\phi_1,\phi_2,\ldots\}$ is a complete Schauder basis for $ X$ and because $P<\infty$, there exists a number $N$ such that for all $n>N$ and for all $p$, $1\le p\leq P$, there exists a $\xi_{p,n}\in S_n\cap B_1$ such that 
\begin{equation}\label{e2}
	\|\xi_{p,n}-\xi_{p}\|_{ X} < \frac{\beta-\alpha}{2\|T\|}.
\end{equation}
Therefore, for all $y\in TB_1$ and for all $n>N$ there exists a $p, 1\leq p\leq P$ and a $\xi_{p,n}\in S_n\cap B_1$ such that
\begin{eqnarray}
\|T\xi_{p,n} - y\|_{ Y} &=& \|T\xi_{p,n}-T\xi_p + T\xi_p - y\|_{ Y}\nonumber\\
&\leq& \|T\xi_{p,n}-T\xi_p \|_{ Y}+ \|T\xi_p - y\|_{ Y} \nonumber\\
&<& \|T(\xi_{p,n}-\xi_p)\|_{ Y} + \frac{\beta-\alpha}{2}\nonumber\\
&<& \|T\|\frac{\beta-\alpha}{2\|T\|} + \frac{\beta-\alpha}{2}\nonumber\\
&=& \beta-\alpha \label{e3}.
\end{eqnarray} 
We get the first inequality above from the triangle inequality, the
second one from inequality~\eqref{e1} and the final one from
inequality~\eqref{e2}. From inequality~\eqref{e5} and the definition
of the number of degrees of freedom, we know that for all $n\in
\bbZ^{+}$ there exists a set of vectors
$\{\psi_{1,n},\ldots,\psi_{m-1,n}\}\subset Y$ such that    
\begin{equation}\label{e6}
   y \in \span_\alpha\{\psi_{1,n},\ldots,\psi_{m-1,n}\}
\end{equation}
for all $y\in T(S_{n}\cap B_1)$.

But, from the definition of the number of degrees of freedom and
inequality~\eqref{e4} we know that for all $n\in\bbZ^{+}$ and all sets
of vectors $\{\psi_{1,n},\ldots,\psi_{m-1,n}\}$ there exists a vector $\psi\in TB_1$ such that 
\begin{equation*}
	\psi\notin \span_\beta\{\psi_{1,n},\ldots,\psi_{m-1,n}\}.
\end{equation*}
From inequality~\eqref{e3} we know that for all $n>N$ there exists a $\xi_{p,n}\in S_n\cap B_1$ such that 
\begin{equation*}
	\|T\xi_{p,n}-\psi\| < \beta-\alpha.
\end{equation*}
Therefore, for all $n>N$ there exists a $\xi_{p,n}\in S_n\cap B_1$ such that 
\begin{equation}
	T\xi_{p,n} \notin \span_\alpha\{\psi_{1,n},\ldots,\psi_{m-1,n}\}.
\end{equation}
This directly contradicts condition~\eqref{e6}. Therefore, if $\epsilon_m$ exists then $\epsilon_{m,n}$ exists for $n$ large enough and
\begin{equation*}
	\lim_{n\to\infty} \epsilon_{m,n} = \epsilon_m.
\end{equation*} 
\end{proof}

\emph{Theorem~\ref{CScomSinTh:calc}.}
Let $X,Y$ be two finite dimensional Banach spaces and let $T:X\to Y$
be a linear operator. Also let $B_1$ be the closed unit ball in $X$
and suppose $\Psi_p$ is defined as in Section~\ref{NcomSin}. Then
\begin{equation*} 
\sup_{x\in B_1} \|Tx\|_Y = \epsilon_1
\end{equation*}
and for all $p\in\bbZ^{+}$
\begin{equation*}
\inf_{\{\psi_i\}_{i=1}^p\in\Psi_p}\sup_{x\in B_1}\inf_{a_1,\ldots,a_{p}} \left\|Tx-\sum_{i=1}^p a_i\psi_i\right\|_{Y} = \epsilon_{p+1}.
\end{equation*}
\begin{proof}
Let $\epsilon_{p+1}'$ denote the left hand side of the above equation.
Assume $\epsilon_{p+1}'<\epsilon_{p+1}$. Then there exists a set
$\{\psi_i\}_{i=1}^p\in\Psi_{p}$ such that
\begin{equation*}
\epsilon_{p+1}'':=\sup_{x\in B_1}\inf_{a_{1},\dots,a_{p}}
\left\|Tx-\sum_{i=1}^p a_i\psi_i\right\| < \epsilon_{p+1}. 
\end{equation*}
By definition this implies $\cN(\epsilon_{p+1}'')\leq p$, a
contradiction to $\inf_{\epsilon<\epsilon_{p+1}}\cN(\epsilon)\geq
p+1$. Hence $\epsilon_{p+1}'\geq\epsilon_{p+1}$.
Now assume $\epsilon_{p+1}'>\epsilon_{p+1}$.
Let
$\epsilon\in(\epsilon_{p+1},\epsilon'_{p+1})$. 
From $\sup_{\epsilon>\epsilon_{p+1}}\cN(\epsilon)=p$ it follows
$\cN(\epsilon)\leq p$.
Hence there exists a set $\{\psi_i\}_{i=1}^p\subset Y$ such that 
\begin{equation*}
\sup_{x\in B_1}\inf_{a_{1},\dots,a_{p}} \left\|Tx-\sum_{i=1}^p a_i\psi_i\right\| 
\leq \epsilon<\epsilon'_{p+1}.
\end{equation*}
Therefore $\{\psi_i\}_{i=1}^p \in \Psi_{p,\epsilon} \subset \Psi_p$ and
\begin{equation*}
\epsilon_{p+1}'>\inf_{\{\psi_i\}_{i=1}^p\in\Psi_p}\sup_{x\in B_1}\inf_{a_{1},\dots,a_{p}} 
\left\|Tx-\sum_{i=1}^p a_i\psi_i\right\|,
\end{equation*} 
a contradiction. Hence $\epsilon_{p+1}'=\epsilon_{p+1}$.
\end{proof}
\emph{Theorem~\ref{Nthe:comp}}. (Converse to Theorem~\ref{Nthe:defBan})
Suppose $X$ and $ Y$ are normed spaces with norms $\|\cdot\|_{ X}$ and
\mbox{$\|\cdot\|_{ Y}$}, respectively, and $T: X\to  Y$ is a bounded linear
operator. If for all $\epsilon > 0$ there exist $N\in \bbZ_0^+$ and
a set $\{\psi_i\}_{i=1}^N\subset  Y$ such that for all $x\in \bar{B}_{1,X}(0)$
\begin{equation*}
    \inf_{a_1,\ldots,a_N} \left\|Tx-\sum_{i=1}^N a_i\psi_i\right\|_{ Y}
    \leq \epsilon
\end{equation*}
then $T$ is compact.

\begin{proof} 
We prove that $T$ is compact by showing that the set
$T(\bar{B}_{1,X}(0))$ is totally bounded. Let $\delta > 0$ be
given. Then there exist an $N\in \bbZ_0^+$ and a set
$\{\psi_i\}_{i=1}^N\subset  Y$ such that for all $x\in \bar{B}_{1,X}(0)$
\begin{equation} \label{eqa0}
    \inf_{a_1,\ldots,a_N} \left\|Tx-\sum_{i=1}^N a_i\psi_i\right\|_{ Y}
    \leq \frac{\delta}{4}.
\end{equation}
For any given $x \in \bar{B}_{1,X}(0)$ we can choose
${a}_i^{x}$, $i = 1,\ldots,N$ such that  
\begin{eqnarray} 
   && \left\|Tx-\sum_{i=1}^N {a}_i^{x}\psi_i\right\|_{ Y} \nonumber \\ &\leq& \inf_{a_1,\ldots,a_N} \left\|Tx-\sum_{i=1}^N a_i\psi_i\right\|_{ Y} + \frac{\delta}{4} \label{eq43} \\
    &\leq& \frac{\delta}{2}\label{ez23}.
\end{eqnarray} 
Here, the last inequality follows from~\eqref{eqa0}.
Also, because we can choose $a_i = 0$ for $i = 1,\ldots,N$, for all $x\in \bar{B}_{1,X}(0)$
\begin{equation} \label{eqal1}
    \inf_{a_1,\ldots,a_N} \left\|Tx-\sum_{i=1}^N a_i\psi_i\right\|_{ Y}
    \leq \|Tx\|_Y.
\end{equation}
Substituting inequality~\eqref{eqal1} into~\eqref{eq43} and using the triangle inequality, we get
\begin{equation} \label{eqak5}
    \left\|\sum_{i=1}^N {a}_i^{x}\psi_i\right\|_{ Y}
    \leq 2\|Tx\|_Y + \frac{\delta}{4} \leq 2\|T\| + \frac{\delta}{4}.
\end{equation}
We get the last inequality from the boundedness of $T$. Because the
span of $\psi_1,\ldots,\psi_N$ is finite dimensional and because of
the uniform bound~\eqref{eqak5}, there exists a
finite set of elements $\{y_1,\ldots,y_M\}\subset Y$ such that for all
$x\in \bar{B}_{1,X}(0)$
\begin{equation}\label{ez1}
\inf_{i=1,\ldots,M} \left\|y_i - \sum_{j = 1}^N {a}_j^{x}\psi_j\right\|_Y \leq \frac{\delta}{2}.
\end{equation} 
From inequalities~\eqref{ez1} and~\eqref{ez23} and the triangle inequality we get 
for all $x\in\bar{B}_{1,X}(0)$
\begin{equation}
\inf_{i=1,\ldots,M} \|y_i - Tx\|_Y \leq \delta.
\end{equation} 
Therefore, the $y_i$, $i=1,\ldots,M$ form a finite $\delta$-net for $T(\bar{B}_{1,X}(0))$ and therefore $T(\bar{B}_{1,X}(0))$ is totally bounded. Hence, $T$ is compact.
\end{proof}

%%% Local Variables: 
%%% mode: latex
%%% TeX-master: "main"
%%% End: 

\bibliography{ref,IEEEabrv}

%\begin{biography}
%[{\includegraphics[width=1in,height=1.25in,clip,keepaspectratio]{./fig/jochen}}]
%{Jochen Trumpf}
%(M'04) received the Dipl.-Math. and Dr. rer. nat. degrees in mathematics from
%the University of W{\"u}rzburg, Germany, in 1997 and 2002, respectively.
%He is working as a Research Fellow for the Department of Information
%Engineering in the Research School of Information Sciences and Engineering at
%The Australian National University, Canberra, Australia, and is currently
%seconded to National ICT Australia Ltd.
%His research interests include observer theory and design, linear systems
%theory and optimisation problems in digital communication.
%\end{biography}
%\begin{biography}
%[{\includegraphics[width=1in,height=1.25in,clip,keepaspectratio]{./fig/ram}}]
%{Ram Somaraju} received B.Sc. and B.E. degrees in Physics and Computer Systems engineering from
%the University of Auckland, New Zealand, in 2001 and 2003, respectively.
%He is currently pursuing a Ph.D. degree in the Department of Information
%Engineering in the Research School of Information Sciences and Engineering at
%The Australian National University, Canberra, Australia.
%His research interests include underwater radio communications, MIMO communication
%and uncertainty principles for communication channels.
%\end{biography}
\end{document}